\documentclass[9pt,twocolumn,twoside]{osajnl}

\journal{ol} 

\setboolean{shortarticle}{true}

\title{Non-line-of-sight imaging in the presence of scattering media using Phasor Fields}

\author[1*]{Pablo Luesia}
\author[1*]{Miguel Crespo}
\author[1]{Adrian Jarabo}
\author[1]{Albert Redo-Sanchez}

\affil[1]{Universidad de Zaragoza, I3A}

\affil[*]{Joint first authors} 

\ociscodes{(010.1350) Backscattering; (150.0155) Machine vision optics; (110.1758) Computational imaging.}

\usepackage{graphicx}
\usepackage{subcaption}
\usepackage{soul}
\usepackage{booktabs}
\usepackage{multirow}
\usepackage[normalem]{ulem} 
\usepackage{tikz}
\usetikzlibrary{tikzmark}

\newcommand{\new}[1]{{#1}}

\newcommand{\tableFig}[2]{\raisebox{-.5\height}{\includegraphics[width=#2]{#1}}}
\newcolumntype{C}{>{\centering\arraybackslash}m{7em}}
\newcolumntype{\px}{>{\centering\arraybackslash}m{7em}}

\newcommand{\px}{\mathbf{x}}

\begin{abstract}
  Non-line-of-sight (NLOS) imaging aims to reconstruct \new{partially or completely occluded scenes}. Recent approaches have demonstrated high-quality reconstructions of complex scenes with arbitrary reflectance, occlusions, and significant multi-path \new{effects}. However, previous works focused on surface scattering only, which reduces its generality in more challenging scenarios such as scenes submerged in scattering media. In this work, we investigate \new{current state-of-the-art NLOS imaging methods based on \emph{Phasor Fields} to reconstruct scenes submerged in scattering media}. We empirically analyze the \new{capability} of \emph{Phasor Fields} in \new{reconstructing complex} synthetic scenes \new{submerged in thick} scattering media. We also apply the method on real scenes, showing that it performs similarly as recent diffuse optical tomography methods.
\end{abstract}

\setboolean{displaycopyright}{true}

\begin{document}

\maketitle


Recent advances in transient imaging~\cite{jarabo2017} using ultrafast sensors opened a wide range of novel imaging modalities, including the ability of imaging fully or partially occluded scenes, or \emph{non-line-of-sight (NLOS) imaging}~\cite{velten2012,otoole2018, liu2019,lindell2019,xin2019,iseringhausen2020}. NLOS imaging has a large body of applications, including medical imaging, autonomous driving, and surveillance and security, among others.

\begin{figure}
  \centering
  \includegraphics[width=\columnwidth]{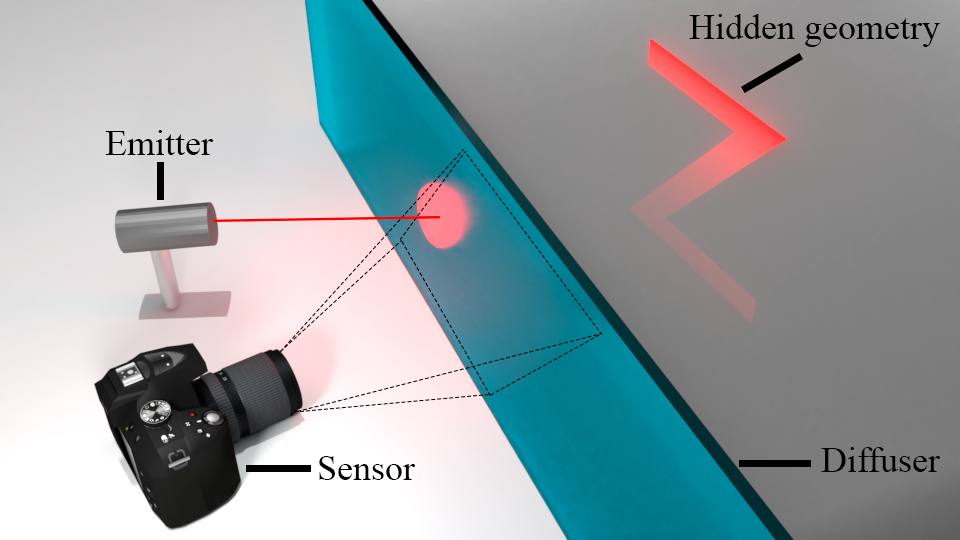}
  \caption{\textbf{Scene setup.} The scene is hidden behind a diffuser, with the scene being submerged in a scattering medium.}\label{fig:teaser_setup}
\end{figure}

NLOS imaging aims to recover hidden scenes by using the light scattered from a visible secondary surface, or \emph{relay wall}~\cite{maeda2019}. Most previous NLOS imaging approaches are based on time-resolved information and filtered backprojection using heuristic filters~\cite{velten2012,laurenzis2014,arellano2017}, or inverting simplistic ad-hoc light transport models~\cite{otoole2018,xin2019,heide2019,iseringhausen2020}. Unfortunately, these approaches do not properly deal with the challenges resulting from multiple scattering, anisotropic reflectance, occlusions, and clutter in the hidden scene. Recent wave-based methods for NLOS imaging~\cite{liu2019,lindell2019,liu2020,nam2021low} have overcome some of these limitations by posing NLOS imaging as a \emph{virtual} wave-propagation problem, \new{which effectively transforms the problem} into a \emph{virtual} line-of-sight (LOS) \new{one}. These \new{methods enable} NLOS imaging of real-world complex scenes at the meter scale.

\new{All these techniques assume that light travels in the vacuum with no scattering media and it is only scattered by the \emph{relay wall}.} This assumption \new{hinders} the applicability of NLOS imaging to scenes in which the presence of scattering media (e.g., smoke or fog) might be significant. Imaging through scattering media is challenging even in LOS setups~\cite{ramachandran1999imaging}, where many techniques have been proposed \new{such as} optical coherence tomography (OCT)~\cite{huang1991,outer1993,han2000, gibson2005}, diffuse optical tomography (DOT)~\cite{yodh1995spectroscopy,o1996imaging,lindell2020,lyons2019}, time-of-flight (ToF) measurements~\cite{heide2014,tobin2019three},
coherent light modulation using spatial light modulators~\cite{vellekoop2007}, or polarization~\cite{demos1997,fade2014long,wu2018}.

In this letter, we empirically analyze the performance of NLOS imaging in scenes \new{submerged in} scattering media (Figure~\ref{fig:teaser_setup}). Given the amount of incoherent light due to scattering, we cannot rely on traditional inversion methods for NLOS imaging. Instead, we leverage recent advances on wave-based NLOS imaging, and in particular on the \emph{Phasor Fields} framework proposed by Liu and colleagues~\cite{liu2019}. \emph{Phasor Fields} enables creating virtual light sources and sensors on the visible surfaces, resulting in robust reconstruction of multiple incoherent scattered light sources~\cite{reza2019, dove2019, teichman2019}. We demonstrate reconstructions of NLOS scenes \new{submerged in} scattering media of increasing density, showing that \emph{Phasor Fields} is \new{capable of} reconstructing scenes in very challenging visibility conditions using a single-frequency laser grid for illumination \new{projected} on a single plane. We hope this work will \new{expand the capability of} NLOS imaging to see through scattering media and foster new avenues of work in the NLOS imaging field.

\new{NLOS imaging using \emph{Phasor Fields} framework~\cite{liu2019} models the virtual light signal by convolving a time-resolved optical carrier with a monochromatic phasor, which allows modeling the propagation of light using a Rayleigh-Sommerfeld diffraction (RSD) operator selecting a suitable wavelength. Given that the spatial separations of the different illumination and sensor points is in the order of centimeters, suitable wavelengths lie in the optical range, whereas longer wavelengths (e.g., radiofrequency) have neither the spatial nor the time resolution to distinguish the different pulses in time.}
The immediate consequence of this observation is that we can use well-known tools from Fourier optics to model sophisticated virtual imaging systems. Therefore, by measuring the impulse light transport matrix of the hidden scene \( H(\px_p \rightarrow \px_c, t) \), we can pose the NLOS imaging problem as a \emph{virtual} LOS one~\cite{reza2019phasor}. Let us define the phasor \( \mathcal{P}_{\omega}(\px, t) \) at point \( \px \), time \( t \), and frequency \( \omega \) as 

\begin{equation}
  \mathcal{P}_{\omega}(\px, t) \equiv \mathcal{P}_{0,\omega}(\px) e^{i \omega t},
\end{equation}

with \( \mathcal{P}_{0,\omega}(\px) \) and \( e^{i \omega t} \) being the intensity and phase at the instant \( t \) of the phasor. We can model the propagation from a surface \( S \) to a point \( \px_d \in D \) of this virtual wave field \( \mathcal{P}_{\omega}(\px, t) \) as

\begin{equation}
  \mathcal{P}_\omega({\px}_d, t) = \gamma \int_{S} {\mathcal{P}_\omega({\px}_s, t)\frac{e^{ik|{\px}_d - {\px}_s|}}{|{\px}_d - {\px}_s|} d{\px}_s}, \label{eq:rsd}
\end{equation}

where \( \gamma \approx 1 / \mid \left< S \right> - \px_d \mid \) is an attenuation factor, \( k \) is the wavelength number (\( k = 2\pi / \lambda \)), being \(\lambda\) the wavelength of the light, \( \omega \) is the frequency of the monochromatic component, \( \px_s \in S \), and \( \px_d \in D \). \eqref{eq:rsd} has the form of a RSD operator. In the following, we remove the frequency dependence from the phasor and consider only monochromatic phasors. Non-monochromatic phasors can be constructed as the superposition of the monochromatic phasor components.

To image a hidden scene from a virtual camera with the aperture placed at surface \( C \), we first need to compute the phasor field on the virtual sensor \( \mathcal{P}(\px_c, t) \) with \( \px_c \in C \) as a function of a phasor \( \mathcal{P}(\px_p, t) \) in the virtual emitter on surface \( P \). This is computed by leveraging the linearity and time-invariance of light transport and using the impulse response of the hidden scene \( H(\px_p \rightarrow \px_c, t) \) to compute \( \mathcal{P}(\px_c, t) \) as

\begin{equation}
  \mathcal{P}(\px_c, t) =
  \int_P \mathcal{P}(\px_p, t) \star H(\px_p \rightarrow \px_c, t) d\px_p,
\end{equation}

where \( \star \) denotes the convolution operator. Finally, to generate the image of the hidden scene \( I(\px_v) \) as seen from the virtual sensor, with \( \px_v \) the point being reconstructed, an image formation model \( \Phi(\cdot) \) is applied over \( \mathcal{P}(\px_c, t) \) as

\begin{equation}
  I(\px_v) = \Phi\left( \mathcal{P}(\px_c, t) \right).
\end{equation}

The image-formation function \( \Phi(\cdot) \) depends on the type of imaging system (details in~\cite{liu2019}). In our work, we use a virtual time-gated camera by setting the emitter phasor \( \mathcal{P}(\px_p, t) \) and image formation function \( \Phi(\mathcal{P}(\px_c, t)) \) to

\begin{equation}
  \mathcal{P}(\px_p, t) = e^{i \omega (t-\frac{1}{c}\left|\px_v-\px_p\right|)}\,e^{- \frac{(t - t_0-\frac{1}{c}\left|\px_v-\px_p\right|)^{2}}{2 \sigma^{2}}} \qquad\text{and} \label{eq:phasor_emision}
\end{equation}

\begin{equation}
    \Phi(\mathcal{P}(\px_c, t)) = \left| \mathcal{R}_{\px_v}\left(\mathcal{P}(\px_c, -\frac{1}{c}|\px_v-\px_c|)\right) \right|^{2}, \label{eq:imageFormationModel}
\end{equation}

where \( c \) is the speed of light in vacuum, \( \mathcal{R}_{\px_v} \) is the RSD operator from \( \px_v \) to \( \px_c \), and \( \sigma \) is the width of the virtual illumination pulse used to focus on the specific voxel \( \px_v \). Thus, the virtual camera is focused on each particular voxel, while the virtual illumination acts as a single pulsed point source.

While \emph{Phasor Fields} have demonstrated very effective at reconstructing NLOS scenes in standard ambient conditions, little has been investigated in relation to its performance in reconstructing scenes submerged in scattering media, which results in more challenging visibility conditions for the experimental equipment operating in the visible range. In the following, we analyze the performance of \emph{Phasor Fields} in scenes hidden behind a visible diffuser and submerged in scattering media of increasing density (see Figure~\ref{fig:teaser_setup}). \new{In this setup, the diffuser, which transmits the light in and out of the scene, acts as the \emph{relay wall}, as opposed to a traditional NLOS setup in which the \emph{relay wall} reflects the light into the scene, generating a geometry for looking around corners. The setup can be considered as a LOS setup through heterogeneous media in which the diffuser acts as an occluder of the scene.} 

\begin{figure}
  \centering
  \captionsetup[subfigure]{justification=centering}
  \includegraphics[width=.8\columnwidth]{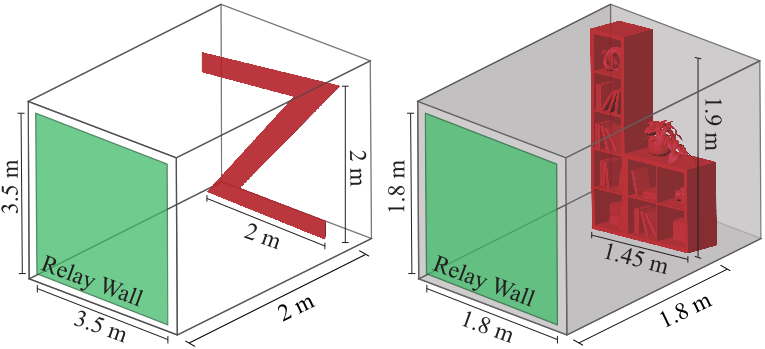}
  \caption{\textbf{Simulated scenes.} We use two simulated scenes: (Left) a single planar letter behind the diffuser (green), and (Right) a closed room with a shelf (red) at the back. }\label{fig:geometry_scenes}
\end{figure}

\newcommand{\zplotsize}{4em}
\begin{figure*}[t]
  \setlength{\tabcolsep}{2pt}
  \graphicspath{{figures/}}
  \captionsetup[subfigure]{justification=centering}
  \vspace{2ex}
  \centering
  \subfloat[][$\mu_t=0$]{\tableFig{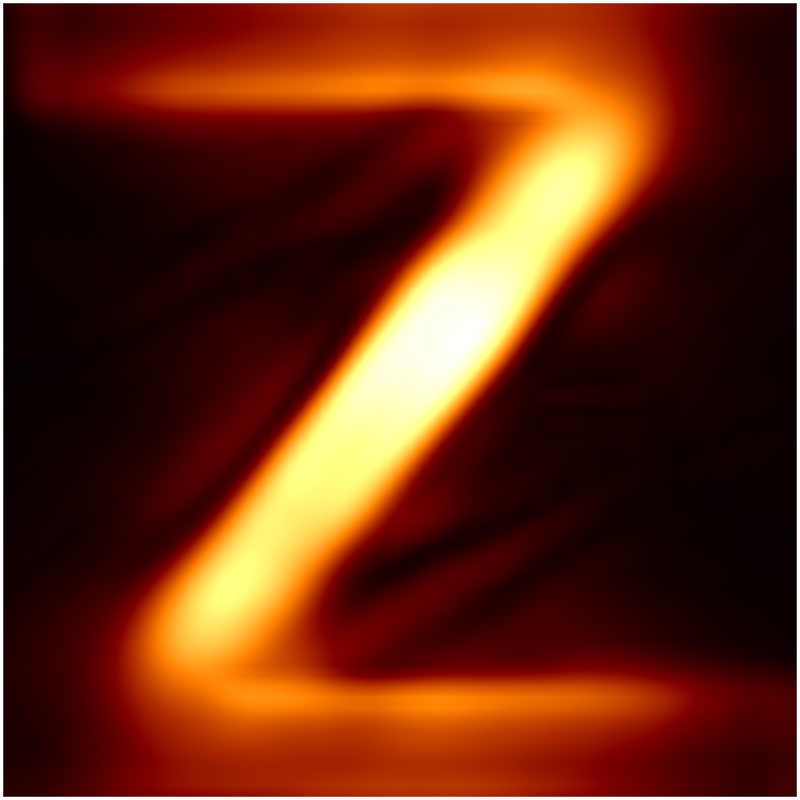}{\zplotsize}\label{fig:z_no_media}}
  \subfloat[][Extinction coefficient $\mu_t$ vs. single scattering albedo $\alpha$]{
  \label{fig:density_results}
      \begin{tabular}{cccccc}
             & \tikzmark{albedo_mu_a} \( \alpha = 0.15 \)& \( \alpha = 0.33\) &
              \( \alpha = 0.5\) & \(\alpha = 0.67\) & 
              \( \alpha = 0.83\) \tikzmark{albedo_mu_b}\\
            \rotatebox[origin=c]{90}{\( \mu_t \) = 0.5} & \tableFig{density/0.5/pf_m1_filtered_atFront.png}{\zplotsize} & \tableFig{density/0.5/pf_m2_filtered_atFront.png}{\zplotsize} & 
            \tableFig{density/0.5/pf_m3_filtered_atFront.png}{\zplotsize} & \tableFig{density/0.5/pf_m4_filtered_atFront.png}{\zplotsize} & 
            \tableFig{density/0.5/pf_m4_filtered_atFront.png}{\zplotsize} \\
            \rotatebox[origin=c]{90}{\( \mu_t \) = 1.0} & \tableFig{density/1/pf_m1_filtered_atFront.png}{\zplotsize} & \tableFig{density/1/pf_m2_filtered_atFront.png}{\zplotsize} & 
            \tableFig{density/1/pf_m3_filtered_atFront.png}{\zplotsize} & \tableFig{density/1/pf_m4_filtered_atFront.png}{\zplotsize} & 
            \tableFig{density/1/pf_m5_filtered_atFront.png}{\zplotsize} \\
            \rotatebox[origin=c]{90}{\( \mu_t \) = 2.0} & \tableFig{density/2/pf_m1_filtered_atFront.png}{\zplotsize} & \tableFig{density/2/pf_m2_filtered_atFront.png}{\zplotsize} & 
            \tableFig{density/2/pf_m3_filtered_atFront.png}{\zplotsize} & \tableFig{density/2/pf_m4_filtered_atFront.png}{\zplotsize} & 
            \tableFig{density/2/pf_m5_filtered_atFront.png}{\zplotsize} \\
      \end{tabular}
      \begin{tikzpicture}[remember picture,overlay]
        \draw[-latex] ([ shift={(0,2ex)}]pic cs:albedo_mu_a) -- node[above] {Decreasing absorption of the media} ([ shift={(0,2ex)}]pic cs:albedo_mu_b);
      \end{tikzpicture}

  }
  \subfloat[][Single scattering albedo $\alpha$ vs. phase function anisotropy $g$ ]{
  \label{fig:anisotropy_results}
      \begin{tabular}{cccccc}
            & \tikzmark{albedo_phase_a} \( \alpha =0.15\) & \( \alpha = 0.33\) & 
            \( \alpha = 0.5\) & \(\alpha = 0.67 \)  & 
            \( \alpha = 0.83\) \tikzmark{albedo_phase_b}\\
         \rotatebox[origin=c]{90}{\( g \) = 0.7} & \tableFig{forwardScattering/pf_m1_filtered_atFront.png}{\zplotsize} & \tableFig{forwardScattering/pf_m2_filtered_atFront.png}{\zplotsize} &
         \tableFig{forwardScattering/pf_m3_filtered_atFront.png}{\zplotsize} & \tableFig{forwardScattering/pf_m4_filtered_atFront.png}{\zplotsize} &
         \tableFig{forwardScattering/pf_m5_filtered_atFront.png}{\zplotsize}\tikzmark{albedo_phase_c}\\
         \rotatebox[origin=c]{90}{\( g \) = 0} & \tableFig{density/1/pf_m1_filtered_atFront.png}{\zplotsize} & \tableFig{density/1/pf_m2_filtered_atFront.png}{\zplotsize} &
         \tableFig{density/1/pf_m3_filtered_atFront.png}{\zplotsize} & \tableFig{density/1/pf_m4_filtered_atFront.png}{\zplotsize} &
         \tableFig{density/1/pf_m5_filtered_atFront.png}{\zplotsize}\\
         \rotatebox[origin=c]{90}{\( g \) = -0.7} & \tableFig{backscattering/pf_m1_filtered_atFront.png}{\zplotsize} & \tableFig{backscattering/pf_m2_filtered_atFront.png}{\zplotsize} & 
         \tableFig{backscattering/pf_m3_filtered_atFront.png}{\zplotsize} & \tableFig{backscattering/pf_m4_filtered_atFront.png}{\zplotsize} & 
         \tableFig{backscattering/pf_m5_filtered_atFront.png}{\zplotsize}\tikzmark{albedo_phase_d}\\
      \end{tabular}
      \begin{tikzpicture}[remember picture,overlay]
        \draw[-latex] ([ shift={(0,2ex)}]pic cs:albedo_phase_a) -- node[above] {Decreasing absorption of the media} ([ shift={(0,2ex)}]pic cs:albedo_phase_b);
        \draw[-latex] ([ shift={(0.3ex,4.4ex)}]pic cs:albedo_phase_c) -- node[above, sloped] {Forward to back scattering} ([ shift={(0.3ex,-4.4ex)}]pic cs:albedo_phase_d);
      \end{tikzpicture}
  }
  \caption{Reconstructions of the \textsc{Z-letter} scene. \subref{fig:z_no_media}) scene with no scattering media. \subref{fig:density_results}) reconstructions with scattering media of varying extinction \( \mu_t \) (in \( m^{-1} \)) and single scattering albedo \( \alpha \). \subref{fig:anisotropy_results}) reconstructions with scattering media of fixed extinction \( \mu_t = 1 \) \( m^{-1} \), varying scattering albedo \( \alpha \), and phase function's anisotropy \( g\). \emph{Phasor Fields} is able to reconstruct the scene even in the presence of highly scattering media.
  }
\end{figure*}

\paragraph{Experimental design. }
We perform our experiments using both simulated data and experimental data. The simulated scenes allow controlling the properties of the scattering media, while the experimental data allows assessing the behavior in real-world scenarios.

\new{For \textbf{simulation,} we use two different scenes (Figure~\ref{fig:geometry_scenes}): A simple scene with a planar Z-letter located at 2 meters away from the diffuser, and a complex scene containing a shelf located 1.8 meters away from the diffuser and completely enclosed in a room (the diffuser is one of the walls of the room). Phasor Fields requires a \emph{relay wall} surface. The diffuser is a planar surface in both scenes that acts as the \emph{relay wall}, which also serves as boundary between the scattering media and the outside world. We avoid undesired reflections, being the only mismatch between the Fresnel index at the \emph{relay wall}. The \textsc{Z-letter} scene allows studying how the medium affects the reconstructions. The \textsc{shelf} scene is more challenging due to feature occlusions and multi-path effects and it is useful to validate the results derived from the \textsc{Z-letter} scene in more complex scenarios.}


We characterize the scattering media by using their bulk optical parameters. In particular, we use the extinction coefficient \( \mu_t = \mu_a + \mu_s \) [in \( m^{-1} \)], and the scattering albedo \( \alpha = \mu_s / \mu_t \) [unitless], with \( \mu_a \) and \( \mu_s \) the absorption and scattering coefficients [in \( m^{-1} \)], respectively. Intuitively, the extinction is related to the medium density and the albedo is related to the strength of the scattering. We model the directionality of the scattering by using the Henyey-Greenstein~\cite{henyey1941} phase function, which models the directionality using an anisotropy factor \( g \in (-1, 1) \), where \( g > 0 \), \( g = 0 \), and \( g < 0 \) for forward, isotropic, and backward scattering, respectively. Unless stated otherwise, we assume an isotropic phase function \( g = 0 \) and homogeneous media.

We use a publicly available transient renderer software~\cite{jarabo2014} to compute the impulse response \( H(\px_p \rightarrow \px_c, t) \) of the hidden scenes. We compute each impulse response for \( 128^2 \) laser positions \( \px_p \) equally spaced in a grid along the visible diffuser, and captured in a single SPAD point \( \px_c \). The temporal resolution is 4096 bins, each with a temporal resolution of about \new{3 ps}. We use the \textbf{experimental} data captured by Lindell and Wetzstein~\cite{lindell2020}, which corresponds to the impulse response of a scene behind a \new{polyurethane foam slab} using lasers and SPAD sensors. They captured \( 32^2 \) equally-spaced positions \( \px_c \) using a confocal configuration (\( \px_c = \px_p \)). \new{The scenes are located behind a 2.54 cm thick slab that acts as both the diffuser and a dense scattering media. The scene itself is not submerged in any scattering medium. The estimated scattering properties values for the slab are \(\mu_t \approx 262.52 \, [\text{m}^{-1}]\) and \(\alpha \approx 0.99\) [unitless]}.


\paragraph{Reconstruction. }
We reconstruct a volumetric representation of the scenes by sequentially focusing the virtual imaging system on each voxel. The voxelization resolution is \( 73 \times 59 \times 73 \) for the \textsc{Z-letter} scene, \( 142^3 \) for the \textsc{shelf} scene, and \( 32^3 \) for all the \textsc{captured} scenes from Lindell and Wetzstein. We use the Matlab solver provided by the authors~\cite{liu2019}. On an Intel Xeon E5 with 256 GB RAM, reconstructions took between 5 to 254 seconds. 


Following Liu et al.~\cite{liu2019}, we set the Gaussian pulse~\eqref{eq:phasor_emision} with a central wavelength \( \lambda = 4 \, \Delta_c \) with \( \Delta_c \), being the distance between sampled points on the relay wall. The pulse has a complete width of \( 4\lambda\) in the simulated data, and \(8\lambda\) in the experimental data ( \( \sigma = 4 \lambda / 6 \) and \( \sigma = 8\lambda/6 \) respectively). Similar to LOS Fourier optics, the wavelength defines the maximum resolution of the reconstruction. Later, we analyze the effect of \( \lambda \) in depth of the reconstructions. For visualization, we remove the contribution of the extinction of the medium by filtering the final reconstruction in post-process. We exploit the spatial information of the voxels, as well as the known properties of the medium, and apply the following scale factor \( K(\px_v) \) for each voxel

\begin{equation}
  K(\px_v) = ( 1 - \alpha e^{-|\px_v - \px_c|\, \alpha }) / ( e^{-|\px_v - \px_c|\,\mu_t} ), \label{eq:PostProcessingFilter}
\end{equation}

which depends on the bulk optical properties of the medium. On a real-world scenario, the media properties are usually unknown, but could be obtained from the measurements using inverse scattering~\cite{gkioulekas2016evaluation}.

\paragraph{Analysis.}
Figure~\ref{fig:density_results} shows the results of \emph{Phasor Fields} when reconstructing the \textsc{Z-letter} scene as a function of different medium parameters (the extinction coefficient \( \mu_t \), and scattering albedo \( \alpha \)). For low scattering albedo, as the medium thickness increases, the amplitude of the signal is attenuated, although the coherence remains and, hence, it is still possible to reconstruct a sharp image. In contrast, a higher scattering albedo results into less coherence and, therefore, into the loss of high-frequency information. 

Figure~\ref{fig:anisotropy_results} shows the effect on the reconstructions of the phase function's anisotropy. For forward scattering, the phasor penetrates deeper into the medium, whereas backward scattering results into more scattering near the virtual camera and, therefore, poorer visibility conditions.

\begin{figure}[ht]
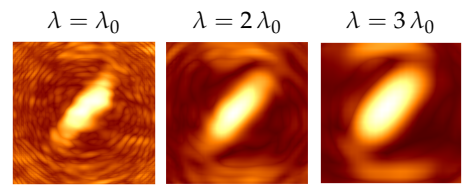

  \setlength{\tabcolsep}{2pt}
  \graphicspath{{figures/}}
  \centering
  \begin{tabular}{ccc}
     \( \lambda = \lambda_0 \) & \( \lambda = 2\,\lambda_0 \) & \( \lambda = 3\,\lambda_0 \) \\
      \tableFig{density/1/pf_m5_filtered_atFront.png}{6em} & \tableFig{lambdax2/1/pf_m5_filtered_atFront.png}{6em} & \tableFig{lambdax4/1/pf_m5_filtered_atFront.png}{6em} \\
  \end{tabular}
  \caption{Reconstructions of the \textsc{Z-letter} scene in the presence of a scattering medium (\( \mu_t = 1 m^{-1} \) and \( \alpha = 0.83 \) ) for increasing wavelength \( \lambda \), with baseline \( \lambda_0 = 4\Delta_c \) and \( \Delta_c = 0.11 m \). Higher values of \( \lambda \) result into deeper penetration through the medium, at the cost of lower spatial.}\label{fig:wavelength_results}
\end{figure}

Now we analyze the effect of the phasor wavelength \( \lambda \) on the reconstruction depth. Figure~\ref{fig:wavelength_results} compares the reconstruction in a highly scattering medium (\( \mu_t \) = 1 \( m^{-1} \), \( \alpha = 0.83 \)) for increasing \( \lambda \). In particular, we double and triple the optimal central wavelength defined by \( \lambda = 4 \Delta_c \). This results into a resolution loss, visible as smoother reconstructions. However, we still can recover the geometry hidden by the scattering medium.

We also analyze how our experiments extrapolate to scenes with significantly more complexity. Figure~\ref{fig:shelves_results} shows the NLOS reconstruction of the \textsc{shelf} scene. The scene contains surface-to-surface multi-path, and multiple scattering. This results in larger indirect contribution to the medium, thus resulting in more challenging visibility conditions. As our results show, even for a relatively thick medium, we can reconstruct the scene with good quality.

\newcommand{\sizeShelf}{6em}
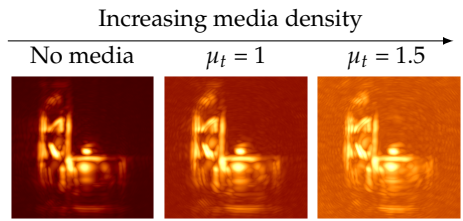
\begin{figure}[t]
  \vspace{2ex}
  \setlength{\tabcolsep}{2pt}
  \graphicspath{{figures/}}
  \centering
  \begin{tabular}{ccc}
    \tikzmark{shelf_a}No media & \( \mu_t \) = 1 & \( \mu_t \) = 1.5 \tikzmark{shelf_b}\\
     \tableFig{shelves/pf_m4_atFront.png}{\sizeShelf}  & 
     \tableFig{shelves/pf_m1_filtered_atFront.png}{\sizeShelf} & 
     \tableFig{shelves/pf_m2_filtered_atFront.png}{\sizeShelf} \\
  \end{tabular}
  \begin{tikzpicture}[remember picture,overlay]
    \draw[-latex] ([ shift={(-2ex,2ex)}]pic cs:shelf_a) -- node[above] {Increasing media density} ([ shift={(2ex,2ex)}]pic cs:shelf_b);
  \end{tikzpicture}
  \caption{Reconstruction of the \textsc{shelf} scene submerged in a medium of increasing density: \( \mu_t = 0 \) (no media), \( \mu_t = 1 \), and \( \mu_t = 1.5 \). In all cases, the scattering albedo is \( \alpha = 0.5 \).}\label{fig:shelves_results}
\end{figure}

\newcommand{\sizeLindell}{6em}
\begin{figure}[h]
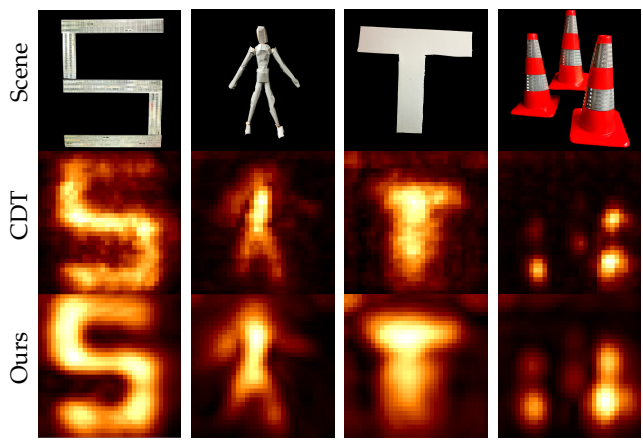

    \setlength{\tabcolsep}{2pt}
    \graphicspath{{figures/}}
    \centering
    \begin{tabular}{ccccc}
        \rotatebox[origin=c]{90}{Scene} & 
        \tableFig{Lindell_PF/letter_s_scene.png}{\sizeLindell} & 
        \tableFig{Lindell_PF/mannequin_scene.png}{\sizeLindell} & 
        \tableFig{Lindell_PF/diffuse_letter_t_scene.png}{\sizeLindell} &
        \tableFig{Lindell_PF/cones_scene.png}{\sizeLindell} \\
        \rotatebox[origin=c]{90}{CDT} &
        \tableFig{Lindell_PF/letter_s_lindell_reconstruction.png}{\sizeLindell} &
        \tableFig{Lindell_PF/mannequin_lindell_reconstruction.png}{\sizeLindell} &
        \tableFig{Lindell_PF/diffuse_letter_t_lindell_reconstruction.png}{\sizeLindell} &
        \tableFig{Lindell_PF/cones_lindell_reconstruction.png}{\sizeLindell} \\
        \rotatebox[origin=c]{90}{Ours} & 
        \tableFig{Lindell_PF/letter_s_reconstruction.png}{\sizeLindell} &
        \tableFig{Lindell_PF/mannequin_reconstruction.png}{\sizeLindell} &
        \tableFig{Lindell_PF/diffuse_letter_t_reconstruction.png}{\sizeLindell} &
        \tableFig{Lindell_PF/cones_reconstruction.png}{\sizeLindell} \\
    \end{tabular}
    \caption{ Reconstructions of experimental data. Each column is a different scene. Top: images of the scenes behind the diffuser. Middle: Lindell and Wetzstein reconstructions~\cite{lindell2020} (CDT). Bottom: our reconstructions using \emph{Phasor Fields}. The higher-frequency of CDT is related to the deconvolution to compensate scattering at the diffuser.
    }\label{fig:real_world_results}
\end{figure}

Finally, we analyze the performance of \emph{Phasor Fields} on the confocal data captured by Lindell and Wetzstein~\cite{lindell2020}. Similar to ours, their method is based on an inversion technique used for NLOS imaging (f-k migration~\cite{lindell2019}). The qualitative results shown in Figure~\ref{fig:real_world_results} are on-par with Lindell and Wetzstein's method, without any modification of the \emph{Phasor Fields} framework, demonstrating its flexibility even in challenging conditions.

In conclusion, we have empirically analyzed the performance of \emph{Phasor Fields} in the presence of scattering media. We have leveraged the capability of \emph{Phasor Fields} in transforming a NLOS problem into a virtual LOS one, in which the scattering media acts as an additional occluder. Our empirical analysis shows that it is feasible to reconstruct NLOS scenes in very challenging visibility conditions under a wide range of media properties. Moreover, we have also shown that, by increasing the reconstruction wavelength, we can partially remove the contribution of the medium at the cost of spatial resolution. However, our work is limited by the same challenges present in LOS systems and, therefore, we can expect that off-the-shelf \emph{Phasor Fields} for very dense scattering media will not be optimal at reconstructing a hidden scene.  However, given the flexibility of \emph{Phasor Fields} for developing virtual complex optical systems, a possible avenue of work is to translate current LOS methods for vision through scattering media into \emph{Phasor fields}.


\paragraph{Funding. }
European Research Council (682080); Defense Advanced Research Projects Agency (HR0011--16-C-0025).

\paragraph{Disclosures. }
The authors declare no conflicts of interest.

\paragraph{Data Availability Statement.}
Data underlying the results presented in this paper are not publicly available at this time but may be obtained from the authors upon reasonable request.


\bibliography{bibliography}

\bibliographyfullrefs{bibliography}

\end{document}